# Enhancing Early Lung Cancer Detection on Chest Radiographs with AI-assistance: A Multi-Reader Study

Gaetan Dissez,[1,++] Nicole Tay,[1,++] Tom Dyer,[1,++,★] Matthew Tam,[2] Richard Dittrich,[3] David Doyne,[4] James Hoare,[4,5] Jackson J. Pat,[3] Stephanie Patterson,[4] Amanda Stockham,[4] Qaiser Malik,[1] Tom Naunton Morgan,[1] Paul Williams,[1] Liliana Garcia-Mondragon,[1] Jordan Smith,[1] George Pearse,[1] Simon Rasalingham[1]

[1] Behold.ai, 180 Borough High St, London SE1 1LB, UK
[2] Cleveland Clinic London Hospital, 33 Grosvenor Place, London SW1X 7HY, UK
[3] Mid South Essex NHS Foundation Trust, Prittlewell Chase, Westcliff-on-Sea, SS0 0RY, UK
[4] King's College Hospital NHS Foundation Trust, Denmark Hill, London SE5 9RS, UK
[5] London South Bank University, 103 Borough Road, London SE1 0AA, UK

Submitted: 31st August 2022
[++] equally contributing authors

**Abstract**

**Objectives** The present study evaluated the impact of a commercially available explainable AI algorithm in augmenting the ability of clinicians to identify lung cancer on chest X-rays (CXR).

**Design** This retrospective study evaluated the performance of 11 clinicians for detecting lung cancer from chest radiographs, with and without assistance from a commercially available AI algorithm (red dot®, Behold.ai) that predicts suspected lung cancer from CXRs. Clinician performance was evaluated against clinically confirmed diagnoses.

**Setting** The study analysed anonymised patient data from an NHS hospital; the dataset consisted of 400 chest radiographs from adult patients (18 years and above) who had a CXR performed in 2020, with corresponding clinical text reports.

**Participants** A panel of readers consisting of 11 clinicians (consultant radiologists, radiologist trainees and reporting radiographers) participated in this study.

**Main outcome measures** Overall accuracy, sensitivity, specificity and precision for detecting lung cancer on CXRs by clinicians, with and without AI input. Agreement rates between clinicians and performance standard deviation were also evaluated, with and without AI input.

**Results** The use of the AI algorithm by clinicians led to an improved overall performance for lung tumour detection, achieving an overall increase of 17.4% of lung cancers being identified on CXRs which would have otherwise been missed, an overall increase in detection of smaller tumours, a 24% and 13% increased detection of stage 1 and stage 2 lung cancers respectively, and standardisation of clinician performance.

**Conclusions** This study showed great promise in the clinical utility of AI algorithms in improving early lung cancer diagnosis and promoting health equity through overall improvement in reader performances, without impacting downstream imaging resources.

**Key words:** Artificial Intelligence – Lung Cancer Diagnosis

## 1 Introduction

Lung cancer is one of the most common cancers worldwide, accounting for an estimated 1.76 million deaths per year (1). According to Cancer Research UK, the net 5-year survival rate for patients diagnosed with Stage One lung cancer is 56.5%, dropping drastically to only 2.9% for Stage IV (2). While prognosis is strongly associated with stage of diagnosis, 75% of lung cancer cases are detected only in its advanced stages with nodal spread and metastatic disease, due to subtle or absence of symptoms in the initial stages of the disease (3). The British Lung Foundation estimated that trachea, bronchus and lung cancers cost the UK a staggering £45 billion in direct, indirect and intangible costs, including welfare and disability/quality-adjusted life years(4). Considering the high mortality and economic burden associated with lung cancer, it is essential that improvements are made in early lung cancer diagnoses.

In an effort to increase detection of lung cancer, screening programmes using low dose computed tomography (CT) have become increasingly endorsed. This follows the National Lung Screening Trial (NLST) which showed that screening in high-risk individuals using low-dose CT can reduce lung cancer mortality by 20% (5). However, there has been much debate on the cost effectiveness of lung cancer screening (6), especially when considering the increase in over diagnoses (7) and additional pressures on already short-staffed NHS radiology departments (8). Furthermore, several studies have highlighted disparities associated with lung cancer screening programmes due to factors such as screening criteria (9), rurality (10) and access to care (11). While low dose CT screening has shown benefits in lung cancer outcomes, infrastructure and cost constraints means that it is inevitable that CXRs will remain a core modality for lung cancer detection (12), especially in regions of higher deprivation. Per current National Institute for Health and Care Excellence (NICE) lung cancer guidelines, CXR is recommended for initial evaluation in all patients, aside from those aged above 40 years who have unexplained haemoptysis (13). However, given the two-dimensional projection nature of CXRs, certain structures on lung fields may be partially or completely obscured, making detection of lung lesions a challenging task (14). A recent systematic review suggested that around 20% of lung cancers are missed on CXRs (15). With advancements in technologies, the use of artificial intelligence (AI) as an assisted reader has demonstrated improvements in sensitivity on CXRs (16–18), especially in less experienced readers (19).

★ corresponding author: tomd@behold.ai

Although several studies have examined how AI improves the performance of readers of varying levels of experience, their scopes are limited in generalisability, and applicability to the NHS is not guaranteed. Furthermore few studies have considered the impact of AI on workflows and clinical decisions. The present study evaluates the impact of a commercially available explainable AI algorithm on the ability of NHS clinicians to identify lung cancer on CXRs. The performance of 11 radiologists and reporting radiographers was compared against ground-truth labels and clinically confirmed cancer diagnoses.

## 2 Materials and Methods

This retrospective study was approved by the local radiology board. All patient imaging was fully anonymised for the purposes of the study. Ethical approval was not required as review of anonymised images was retrospective and was endorsed by the local Caldicott guardian.

### 2.1 Dataset Curation

An information request was made to an NHS hospital to yield an anonymised list of adult patients (18 years and above) who had a chest radiograph (CXR) acquired in 2020 and their corresponding clinical text reports.

A simple text search was implemented in Python to extract clinical text reports containing words and phrases that are typically associated with potentially malignant CXRs: 'CT', 'Alert', 'Mass', 'Nodule', 'Hilar Enlargement', 'Urgent'. A further manual filter of the remaining reports for exams that required a follow up CT Chest examination resulted in a pool of 650 abnormal CXRs, of which a random sample of 200 exams were selected to be included in the study dataset. An additional random sample of 200 CXRs (where there were no urgent findings stated in the original text report) was taken from the original anonymised list. This resulted in a final dataset of 400 CXRs. The workflow for curating the dataset is depicted in Figure 1. The demographic breakdown of the 400 patients is shown in Table 1.

### 2.2 Labelling Protocol

As part of the study, for each CXR, labels were obtained using 3 different methods (as shown in Figure 2). Ground truthed labels were obtained to assess the classification performance of the AI algorithm. In order to assess the impact of AI on clinician performance, a panel of readers was recruited to review the CXRs (with and without AI output). Finally clinical outcomes were tracked for a portion of CXRs to accurately assess the performance of detection of lung cancer on CXRs.

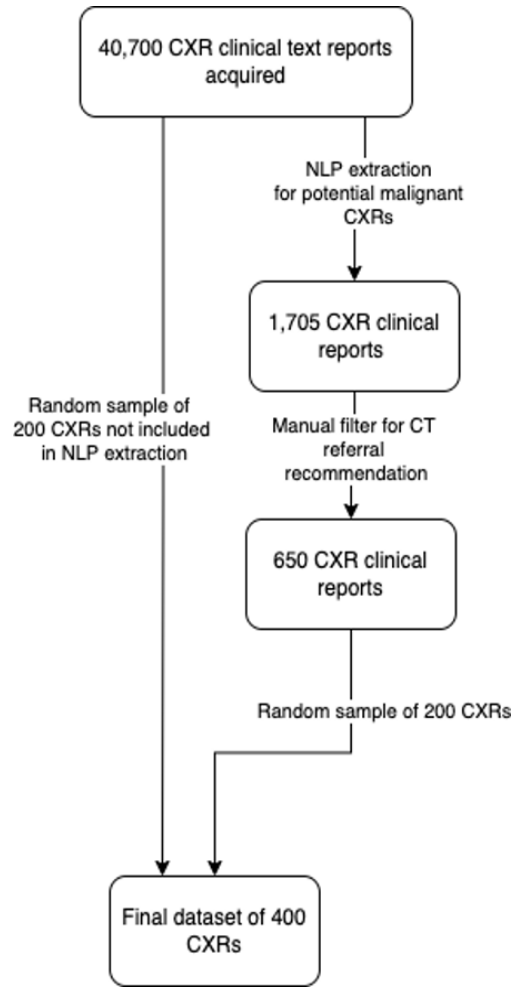

**Figure 1.** Workflow demonstrating the data curation process

#### 2.2.1 Ground Truth

The ground truth for each of the 400 exams was established by individual blind reads of the CXRs by two FRCR consultant radiologists with a minimum of 10 years' experience. In case of a discrepancy, images and conflicting radiologists' labels were sent for arbitration by a third FRCR radiologist. The consultant radiologists had to confirm the absence or presence of visual features of abnormal findings. Any exam that was labelled with the presence of Nodules, Mass or non-vascular Hilar Enlargement were defined as suspicious for lung cancer (SLC). The ground truth indicated that 85% (340/400) of exams in the dataset present with a finding or abnormality (as shown in Table 2). 132 exams (33%) were labelled as SLC according to the ground truth.

#### 2.2.2 Participant Labels

400 patients were selected to evaluate the clinicians' performance at interpreting chest radiographs with and without AI assistance. The participating panel consisted of 11 clinicians: 3 FRCR consultant radiologists, 2 board-certified radiologists, 2 radiology trainees and 4 reporting radiographers. The make-up of the panel is shown in Table 3. The radiologists involved in ground-truthing and providing the clinical outcomes were not included in the panel.

| | Patients | Sex | | Age | | Modality | | View | |
|---|---|---|---|---|---|---|---|---|---|
| | | M/F | %M/F | Mean | Std | DX | %DX | PA | %PA |
| All Patients | 400 | 187/213 | 47%/53% | 68.1 | 16.7 | 80 | 20% | 325 | 81% |
| Confirmed Lung Cancer | 72 | 33/39 | 46%/54% | 72.7 | 11.2 | 8 | 11% | 65 | 90% |
| No Lung Cancer | 328 | 154/174 | 47%/53% | 67.1 | 17.6 | 72 | 22% | 260 | 79% |

Table 1: Patient demographics of the curated dataset

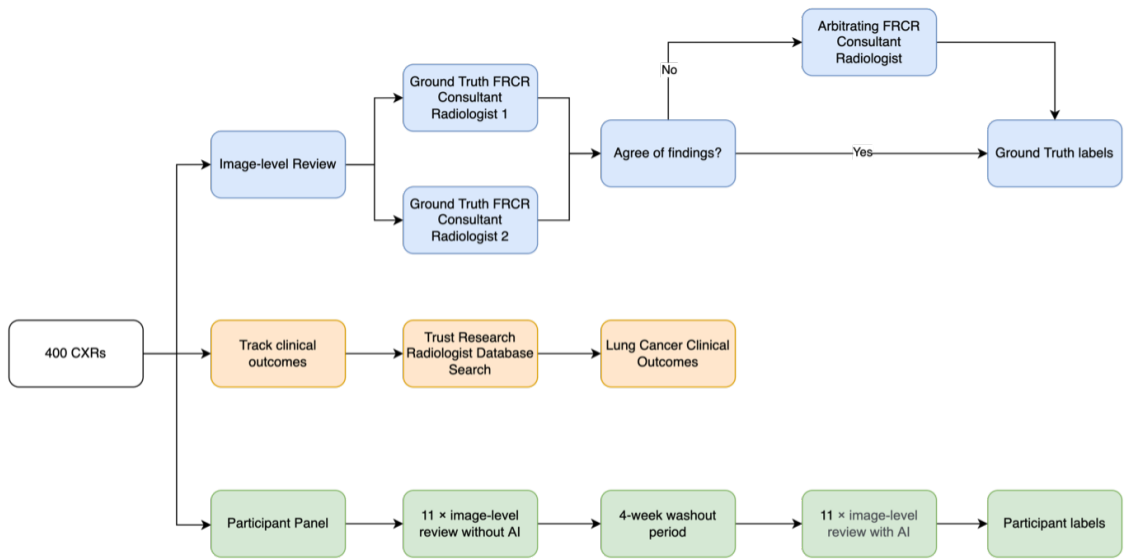

**Figure 2.** Flow diagram describing the labelling protocols used across all 400 CXRs

| Finding | Number of Exams |
|---|---|
| Abnormal | 340 |
| Consolidation | 116 |
| Nodule/Mass | 97 |
| Hilar Enlargement | 62 |
| Pleural Effusion | 33 |
| Pulmonary Oedema | 15 |
| CCF | 11 |
| Atelectasis | 9 |

Table 2: Count of significant findings for the 400 studies, according to the ground-truth established with three FRCR Consultant radiologists

| Participant | Qualification | Years of CXR Reporting Experience |
|---|---|---|
| Participant 1 | Reporting Radiographer | 2 years |
| Participant 2 | Radiology Trainee (pre-FRCR) | 3 years |
| Participant 3 | FRCR Consultant Radiologist | 13 years |
| Participant 4 | Radiology Trainee (pre-FRCR) | 2 years |
| Participant 5 | MD/DNB Radiologist | 18 years |
| Participant 6 | MD/DNB Radiologist | 10 years |
| Participant 7 | FRCR Consultant Radiologist | 11 years |
| Participant 8 | FRCR Consultant Radiologist | 11 years |
| Participant 9 | Reporting Radiographer | 2 years |
| Participant 10 | Reporting Radiographer | 2 years |
| Participant 11 | Reporting Radiographer | 1 year |

Table 3: Participant panel with their associated qualification and years of experience

The study was carried out in two sessions with a washout period of 4 weeks to reduce information bias. Each clinician was asked to independently review the 400 images without AI assistance in the first session, and with AI assistance in the second session. The order in which the images were presented to the clinicians was randomised. In addition to chest radiographs, the clinicians were provided with basic clinical information, including age and sex.

As part of reviewing the chest radiographs, the clinicians were asked to report the absence or presence of abnormalities. If an abnormality is present, they were asked to indicate whether the image had features suggestive of lung cancer (SLC), to mark the location of SLC features and to indicate whether they would recommend follow up imaging (i.e. referral to CT or repeat CXR). Finally, the clinicians were asked to score their confidence level for labelling each image on a 5-point Likert scale from 1: Not at all confident to 5: Completely confident. The images were reviewed using a bespoke web-based tool.

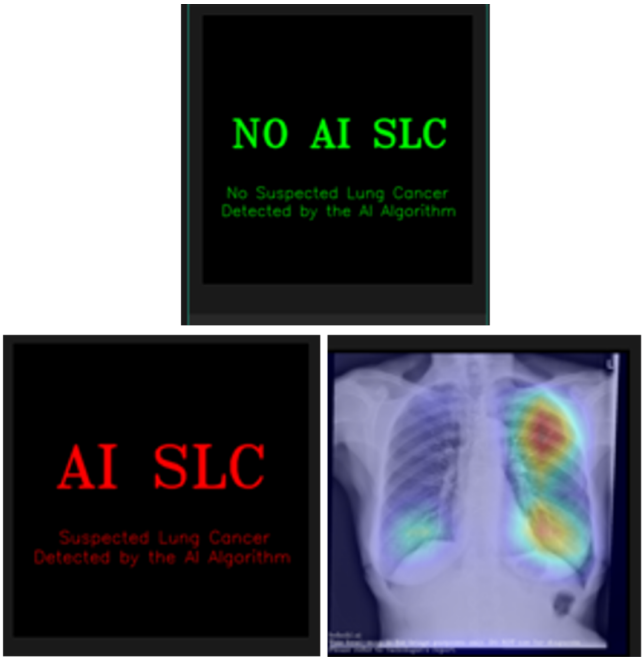

**Figure 3.** Screenshots of AI Output used within the study

Following two sessions of image review, the participating panel were asked to complete a 20-question survey to provide feedback on the study. This included feedback on the study information sheet, training provided and feedback on the AI algorithm outputs. Most questions were multiple choice answers, with an option to provide additional free text responses.

#### 2.2.3 Clinical Outcomes

In addition to the ground truth and participant labels, an FRCR consultant radiologist from the participating Trusts collated downstream clinical outcomes for the 200 CXRs that were originally selected from the abnormal CXR pool. This included information on any follow up imaging including repeat CXR and CT outcomes, lung cancer diagnosis, TNM staging and biopsy outcomes.

### 2.3 AI Solution

AI-based suspected lung cancer prediction was performed by a commercially available AI algorithm (Red Dot®, Behold.ai, London, UK), an ensemble of deep convolutional neural networks. The algorithm triages adult frontal CXRs and will detect findings suspicious for lung cancer (as per ground truth definition). When the AI solution detects an SLC finding, an "AI SLC" notification is shown and a heatmap indicating the algorithm's attention points in produced (as shown in Figure 3). The heatmap was designed to draw radiologists' attention to regions of interest that resulted in the AI output as opposed to a perfect segmentation map. If the AI solution does not identify any SLC finding on the exam, a "No AI SLC" flag is shown. The classification threshold for the algorithmic identification of positive SLC exams was set on an external dataset prior to receipt of the images in the study.

|  | Average Participant without AI [min,max] | AI Model [95% CI] |
|---|---|---|
| Sensitivity | 0.53 [0.21,0.80] | **0.76** [0.68,0.83] |
| Specificity | **0.86** [0.72,0.94] | 0.75 [0.70,0.80] |
| Accuracy | 0.75 [0.69,0.79] | 0.75 [0.71,0.79] |

Table 4: Different metrics for the detection of nodules, masses and hilar enlargements compared to the ground-truth established by FRCR consultant radiologists. The 95% CI stands for 95% Confidence Intervals for the model metrics, using the bootstrap method.

### 2.4 Statistical Analysis

Classification performance of the AI algorithm was assessed against the ground-truthed labels using overall accuracy, sensitivity and specificity. Performance of the 11 clinicians participating in the study (with and without the assistance of AI) was assessed against lung cancer clinical outcomes. Agreement between clinicians was assessed using Cohen's kappa score. A t-test was performed for the comparison of two means, with bonferroni correction (20) applied when multiple tests were conducted. For all analyses, a significance threshold set at <0.05 was applied.

## 3 Results

A total of 400 chest radiographs (53%/47% F/M) were included in the study, of which 132 exams were ground truthed as suspicious for lung cancer (SLC), and 72 were clinically confirmed lung cancers (Table 1).

In this section algorithm standalone performance for detection of suspected lung cancer on a chest radiograph is presented, followed by clinicians' abilities to identify and prioritise patients that are later diagnosed with lung cancer, with and without the AI. Finally, potential biases brought up using AI are explored.

### 3.1 AI standalone performance

When deployed as a standalone algorithm, red dot® identified 167/400 (42%) exams as being suspicious for lung cancer (SLC). The AI algorithm achieved an overall accuracy of 75%, equivalent to the mean performance of the 11 clinicians who participated in the study. As previously demonstrated in our published paper (16), the red dot® algorithm also showed superior sensitivity to 10/11 participants, and marginally lower specificity (Table 4).

### 3.2 Participant misses without AI

In round 1 of the study, every participant individually reviewed 400 exams without AI outputs, resulting in a total of 4,400 CXR reads. Without AI, the participant panel missed 687 nodules, masses and hilar enlargements that were ground truthed as SLC. The AI algorithm was able to pick up 62% (428/687) of those misses. This is in line with our previously published results in (16), which demonstrated that around 60% of missed findings can be detected by the model.

### 3.3 Performance Against Clinically Confirmed Lung Tumours

To evaluate the impact of red dot® on participants' lung tumour detection abilities, the participants' labelling of SLC is compared to the clinically confirmed outcomes. Standalone clinician performances and combined performances where the algorithm is positioned as the first read are presented.

| Metric | Without AI [95% CI] | With AI [95% CI] | Change with AI | Corrected p-value |
|---|---|---|---|---|
| Sensitivity | 0.66 [0.59,0.71] | 0.77 [0.75,0.8] | +0.11 | 0.03 |
| Specificity | 0.81 [0.77,0.85] | 0.75 [0.71,0.77] | -0.06 | 0.17 |
| PPV | 0.44 [0.4,0.48] | 0.41 [0.38,0.43] | -0.04 | 0.58 |

Table 5: Metrics averaged for participants with and without the red dot SLC AI aid. 95% CI stands for 95% Confidence Intervals computed with the bootstrap method.

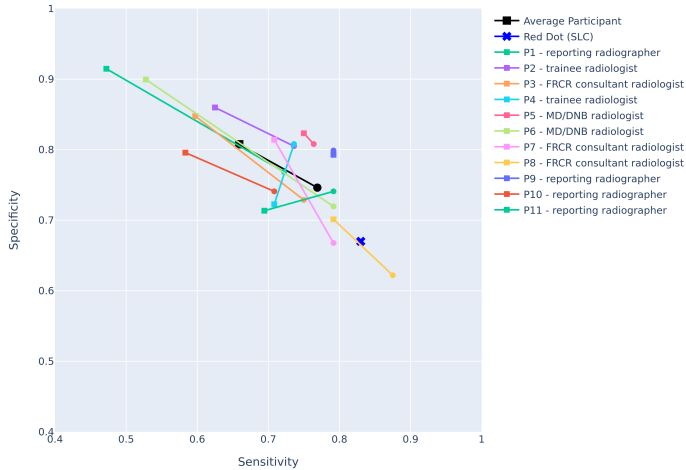

**Figure 4.** Participant's sensitivity and specificity without AI (squares) and with AI (dots). Red dot standalone performance shown with the blue cross, with superior sensitivity to all clinicians. With AI, overall clinician sensitivity increases significantly (as indicated by the black line). With AI, the performance across clinicians is standardised as well, as indicated by the reduce spread of the dots as compared to the squares.

### 3.3.1 Reader Changes with AI

The average performance of 11 clinicians in identifying lung cancer on this dataset is presented in Table 5 With AI, the mean sensitivity of cancer detection increased significantly from 0.66 to 0.77 (+0.11; p-bonferroni=0.03). Mean specificity and precision (PPV) marginally decreased from 0.81 to 0.75 (-0.06) and from 0.44 to 0.41 (-0.04) respectively. Neither decrease was statistically significant (p =0.17 and p =0.58 respectively). Overall changes for sensitivity and specificity when readers are combined with AI are visually represented in Figure 4.

### 3.3.2 Standardisation of the Diagnoses

In addition to improving overall detection of lung tumours, red dot® standardised the overall performance of CXR reporting across the 11 participants. With AI the standard deviation of sensitivity, specificity, and precision decreased by 16%, 14% and 14% respectively (Table 6). As a result, this cohort of lung cancer patients are both more likely to be correctly diagnosed and to receive the same standard of care, independent of who reports their radiograph. This harmonisation can also be seen on Figure 4, where the dots (with AI) are visually more clustered than the squares (without AI), and where the biggest outliers are squares.

|  | Without AI | | With AI | |
|---|---|---|---|---|
|  | Average | Std | Average | Std |
| Sensitivity | 0.66 | 0.10 | **0.77** | **0.04** |
| Specificity | **0.81** | 0.77 | 0.75 | **0.06** |
| PPV | **0.44** | 0.07 | 0.41 | **0.05** |
| Agreement (Cohen-Kappa) | 0.42 | 0.10 | **0.57** | **0.08** |

Table 6: Average Metric Changes with their standard deviation

Furthermore, the standardisation of the diagnosis is reflected in increased agreement between participants. The Cohen-Kappa correlation coefficient increased from 0.42 to 0.57 (+36%) when AI is introduced.

### 3.3.3 Improving Lung Cancer Detection

With the introduction of AI, there was a significant improvement in the detection of confirmed lung tumours, with a sensitivity increase of 0.11 (p = 0.02). This resulted in, on average, 8 more lung cancers identified on CXR with the aid of the AI per participant. Overall, with AI, the number of lung cancers detected increased from 46 to 54, resulting in a 31% reduction in missed lung cancer diagnosis on CXRs.

When considering the total number of individual reads (i.e. 400 CXRs reviewed twice by 11 participants), there were 270 reports which failed to identify lung cancer features on CXRs when AI was not used. With AI, there was a 35% observed reduction in the number of missed lung cancers that were originally labelled incorrectly as normal. In a live clinical setting, it would have been highly likely that such patients would have been discharged, resulting in delayed diagnoses.

Figure 5 shows the 4 exams for which the introduction of AI resulted in positive detection of lung cancer. Every radiograph is shown with the corresponding heatmap generated by the algorithm and provided to the participants. These radiographs contain nodules, a mass, and hilar enlargements that were investigated and later clinically diagnosed as lung tumours.

|  | Without AI | | With AI | |
|---|---|---|---|---|
|  | Average | 95% CI | Average | 95% CI |
| Average Count | 117 | [93,147] | 144 | [119,172] |
| Average Count Lung Cancer | 46 | [38,51] | 54 | [42,59] |
| % Lung Cancer | 39% | - | 38% | - |

Table 7: Patients referred to CT, averaged per participant, when the AI is used and when it is not. 95% CI are 95% Confidence Intervals obtained by the bootstrap method on the average of the 11 participants' outcomes.

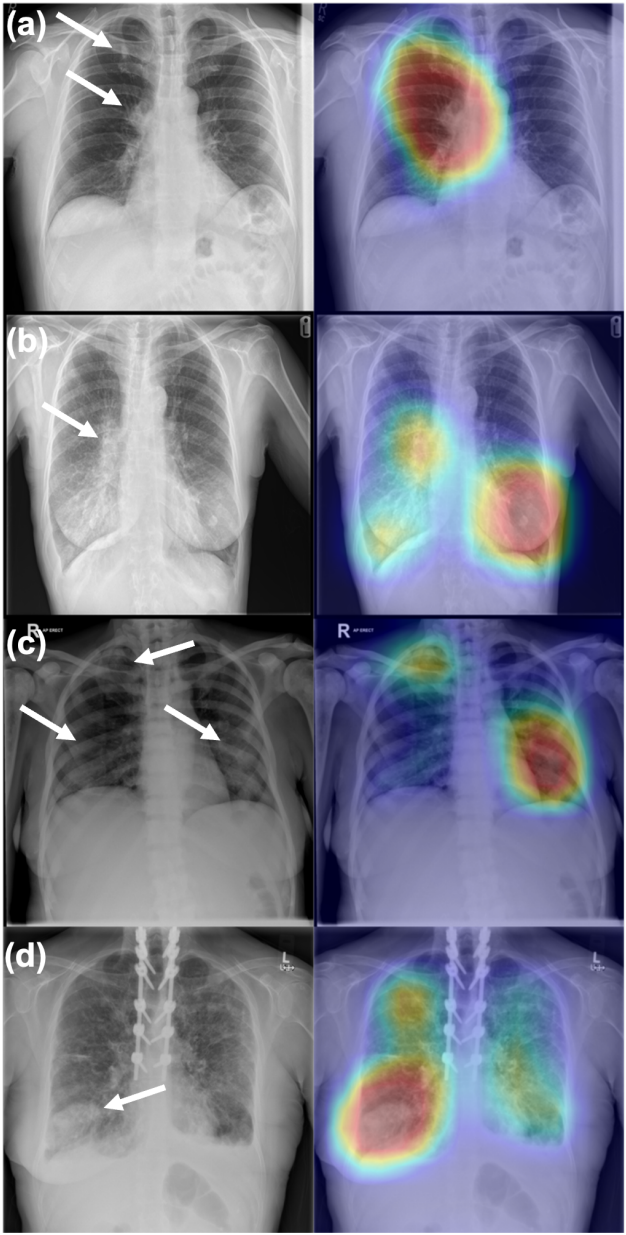

Figure 5. 4 exams that have seen the biggest improvement in lung tumour detection with AI. The white arrows point to the regions of the lungs that show features of the underlying lung cancer. For exams (a), (b), (c), and (d), respectively 9, 6, 5 and 5 more participants reported this exam as SLC with the AI.

### 3.3.4 Impact of workflow

Participants reporting examinations in this study were also required to indicate if they would recommend any follow up imaging, as would be done in a clinical setting. With red dot®, unsurprisingly, the number of patients referred to CT increased from 117 to 144 on average (Table 7). However, the simulated increase in CT referral would have resulted in an increase of 17.4% lung cancers detected, resulting in the proportional change in lung cancer diagnostic CTs being negligible (from 39% to 38%) and statistically non-significant (p= 0.22). An example of a case that was added to the CT referral by 7 participants with the aid of AI is shown in Figure 6. This patient had a CT exam performed in hospital, and lung cancer was excluded owing to infectious findings.

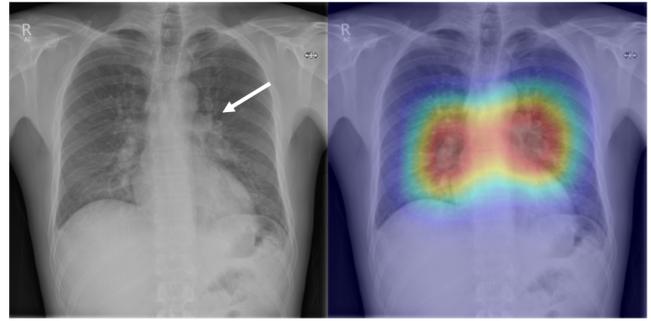

Figure 6. CXR and corresponding heatmap of the patient that saw the biggest CT referral change (+7) with the use of AI. The hospital report also recommends a CT exam because of the left hilum, but a lung cancer diagnosis is later excluded (infection).

### 3.3.5 Tumour size and stage

As shown in Figure 7 and Table 8, participants' sensitivity increased with the use of AI for all sizes and stages of lung cancers. It is generally known that larger tumours and later stage cancers are more easily detectable on a chest radiograph. In this study, we demonstrated that red dot® can improve participants' detection of almost all tumour sizes, including smaller tumours <20mm and early stage cancers as well.

|  | Average Sensitivity | | |
| --- | --- | --- | --- |
|  | **Without AI** | **With AI** | Change |
| Stage I Tumours | 0.52 | **0.64** | +0.12 |
| Stage II Tumours | 0.69 | **0.78** | +0.09 |
| Stage III-IV Tumours | 0.85 | **0.95** | +0.10 |

Table 8: Average sensitivity for all participants on confirmed lung cancers, per stage.

### 3.3.6 Clinician + AI misses

Of the 4,400 individual reads carried out with AI, there were only 15 reports that did not label the exam as SLC due to the red dot® algorithm not detecting features suspicious of lung cancer. Figure 8 shows 4 examples of those exams: all these exams show a large pleural effusion, which were still reported by all radiologists and still recommended for follow up CT. This finding can be suggestive of lung cancer but is not part of the target design of the algorithm, which aims at detecting masses, nodules, and hilar enlargements per the ground truth definitions.

## 3.4 AI and Bias

### 3.4.1 Participant Confidence

As part of the study, participants were asked to assign a confidence level to all reports. Globally, the average confidence

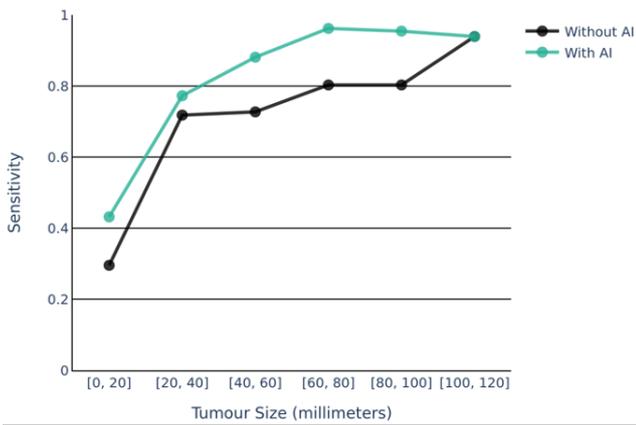

**Figure 7.** Average sensitivity for all participants on confirmed lung cancers, per size

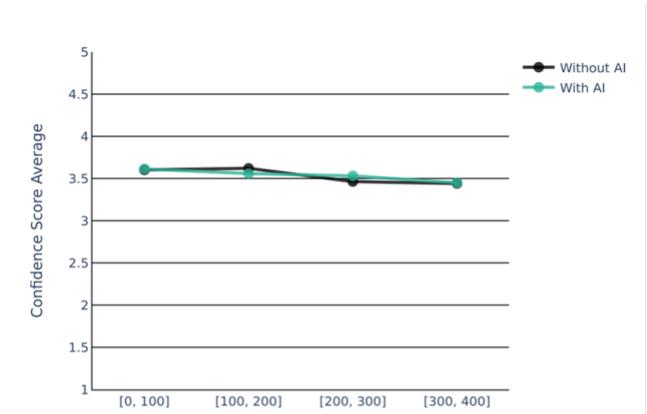

**Figure 9.** Average confidence score over time for the review of the 400 studies, with and without the use of AI.

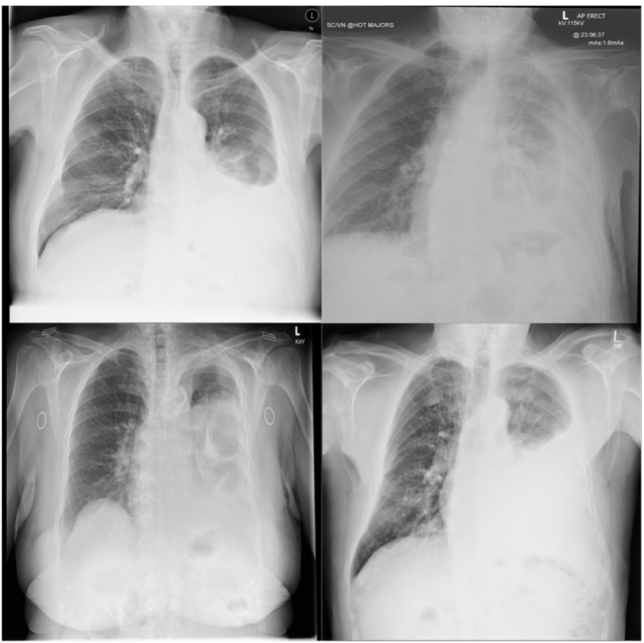

**Figure 8.** 4 examples of radiographs of patients with a confirmed lung cancer that have not been identified by red dot as SLC. These exams have in common that they all present with a large pleural effusion in the left lung.

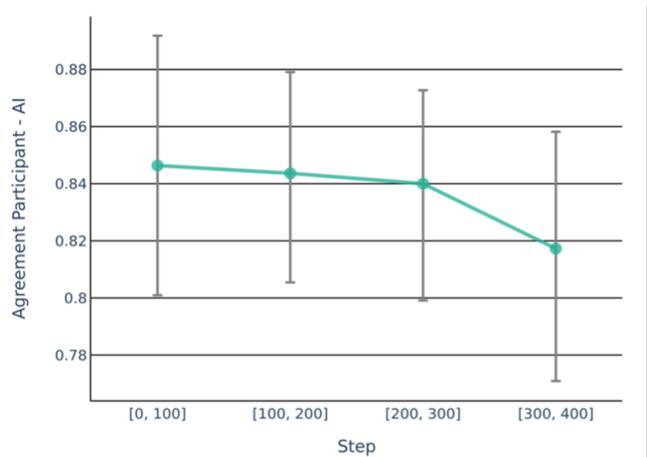

**Figure 10.** Participant's agreement with the model over time. The step corresponds to the order when exams were reported. This order is randomised for every participant, and the agreement is averaged for all participant over a range of 100 exams. The grey ticks correspond to 95% confidence intervals calculated by the bootstrap aggregation method

with and without the AI remained unchanged, with an average confidence of 3.54 in both situations. Further analysis (see Supplementary Material) showed that there was no trend in confidence change, even within participants' subgroups.

To study how reporters' confidence may change over time, participants were presented exams to review in a randomised order. The change of confidence over time is presented Figure 9. On average, with and without the use of AI, participant's confidence decreases marginally when having reviewed more exams. As this trend is also visible when the algorithm is not used, this is likely to be a symptom of a reporter's fatigue as opposed to reliance on the AI output.

### 3.4.2 Participant Agreement with AI

Participant's agreement with red dot® output marginally decreased with the number of radiographs reported over time. This may suggest a decreasing trust in the model's output with the number of exams reported. However, as shown in Figure 10, the high overlap in 95% confidence intervals across the 4 points demonstrates that the decrease in agreement over time is highly variable for participants. Furthermore, the agreement rate with AI output remains significantly higher than the agreement rates between participants. This is likely due to the robust training provided, ensuring that the participants have a clear understanding on the AI outputs.

### 3.4.3 Reporter's Feedback

Upon completion of the study, participants were asked to take part in a survey to collect their feedback on the AI model. 10 of the 11 participants provided their feedback. 8 of the 10 participants declared that reporting was not slower when us-ing the algorithm, and 9 out of 10 reported that the heatmaps produced by the AI model were helpful to understand the algorithm's attention points.

## 4 Discussion

In this retrospective study, the use of an AI solution (red dot®) to triage chest radiographs resulted in increased performance of reporting clinicians to detect lung malignancies and reduced variance in performance across all clinicians. While this is not the first study to evaluate performances across readers of varying levels of expertise (18,21), this is the first study to include reporting radiographers, which is pertinent to the NHS landscape where the reporting radiographer workforce has increased to account for radiology staff shortages (8).

The present analyses reveal that standalone AI algorithm is equivalent to the average performance of 11 reporting clinicians in identifying suspected lung cancer features, with overall accuracy of 75%. This is in line with previously published work where red dot® showed equivalent accuracy but superior sensitivity to consultant radiologists (16). More importantly, the combination of AI and clinician showed improved overall performance for detection of lung tumours, achieving a significant increase in average sensitivity of 11 points, while marginally decreasing specificity and precision. Implementation of AI-based triage resulted in an overall increase of 8 lung cancers (17.4%) being identified on CXRs, which would have otherwise been missed. Additionally, AI assistance resulted in an overall increase in detection of smaller tumours and a 24% and 13% increased detection of stage 1 and stage 2 lung cancers respectively, showing great potential to improve patient survival rates through early detection of lung cancers.

Studies have shown that errors are common in interpretation of chest radiographs due to interobserver variabilities (22–24), resulting in nearly 90% of missed lung cancer cases occurring on CXRs (25). This is the case even for experienced and senior readers, where previous studies have shown significant differences in performance between FRCR consultant radiologists (16). In our study, implementation of AI was shown to standardise performance across 11 reporting clinicians of varying experience levels. The standard deviation of all performance metrics (sensitivity, specificity and precision) decreased when AI was introduced as the first read. Additionally, agreement rate between the clinicians increased by 36%. This would suggest that AI has the potential to reduce inconsistencies across reporting clinicians of all experience levels and improve overall standard of care for patients.

Potential concerns around AI implementation have arisen due to the general perception that AI models are built to avoid misses with a trade-off of increased false positives. The false positives can lead to unnecessary testing (26) and further swamp already stretched imaging services. This study demonstrated that while volume of CT referrals would increase in a simulated workflow when AI is introduced, the proportional increase of 17.4% of lung cancers detected would undoubtedly make this increased imaging worthwhile. Comparatively, low dose CT screening programmes on much larger cohorts, that have been shown to be cost-effective and reduce lung cancer mortality, such as National Lung Screening Trial (NLST) (27) and UKLS (28) have shown 1% and 2% respective pick-up rates of lung cancer.

As AI is increasingly embedded into hospital systems, it is essential to consider susceptibilities in deployment and how it may affect clinical decisions. The greatest benefit of AI in clinical decisions can only be realised when there is a balance of scepticism and trust. Over-reliance on algorithm outputs can lead to medical errors (29–31). The present study showed that overall clinician agreement rate with AI outputs did not increase over time, suggesting that clinicians did not place blind trust on the AI output. This is most likely due to appropriate training, where the participants were instructed to review the chest radiograph first, before viewing the algorithm output (i.e. "AI SLC" flag + associated heatmap, or "No AI SLC" flag). Additionally, the heatmap output allows clinicians to have visual understanding of the algorithm's attention points, thus avoiding the "black box" AI problem [cite Savage N, 2022 Nature]. Feedback from the survey questionnaire indicated that all but one clinician found the heatmap useful (with one saying they "don't know") and all clinicians reporting that they would not change the heatmap output. With the use of red dot® as a triage of CXRs, clinicians are able to make their own clinical judgement on whether they agree with the AI output and make appropriate decisions on whether follow up investigations is required.

Several aspects of the study may not reflect real world clinical settings. It is a limitation of our study that the dataset was collected from a single institution and hence results may not be generalisable, although the participating panel included clinicians from multiple institutions. The participant panel also reviewed the CXRs on personal laptops as opposed to high resolution monitors that would typically be used in a clinical setting. Furthermore, as part of the study design, participants were asked to review the chest radiographs without any additional clinical information (such as reasons for imaging request, patient symptoms and patient history). This allows for a fair comparison between clinician and AI performance, as the algorithm is developed to predict abnormalities on a pixel level only. While there may be other abnormalities on CXRs that can indicate lung malignancy, the red dot® algorithm has only been trained to flag nodules, mass and hilar enlargements as suspicious for lung cancer. Hence there are other radiological findings such as effusion and consolidation that can suggest a lung tumour but are not identified by AI as SLC. However, in a real-world clinical setting, such CXRs will still be reviewed by a clinician and can be referred for further investigation.

The present study demonstrates that the AI algorithm (red dot®) was associated with improved sensitivity for clinical readers on detecting lung cancers on chest radiographs, without negatively impacting downstream imaging resources. The study shows great promise in the clinical utility of the red dot® algorithm in improving early lung cancer diagnosis and promoting health equity through overall improvement in reader performances.

## 5 Footnotes

**Patient and public involvement:** There were no patient and public contributors involved in the design, running and reporting of this study.

**Ethics approval:** Ethics approval was not required for this study. The study was approved by the local radiology board and endorsed by the local Caldicott guardian. All patient imaging was fully anonymised for the purposes of the study.

**Contributors:** GD, NT and TD contributed to conception and design of the work. GD contributed to the analysis and visualisation of the data, drafting of work. NT and TD contributed to interpretation of data, drafting of work, project management. MT contributed to conception of work and acquisition of data. RD, DD, JG, JPP, SP, AS, QM, TNM and PW contributed to reviewing of CXRs within the study. LGM, JS and GP contributed to interpretation of data. All authors revised and reviewed the work critically, gave final approval of the version to be published, and agreed to be accountable for all aspects of the work. NT is responsible for the overall content as the corresponding author.

**Funding:** There is no specific funding for this study.

**Conflicts of interests:** GD, NT, TD, LM, JS, GP, SR are employed by Behold.ai. QM, TNM, PW and SR have stock/stock options in Behold.ai. All other authors have no relevant completing interests to disclose.

**Data availability statement:** Data may be available on request from the corresponding author.

# Supplementary Material

## I. Participants Against Ground Truth Labels

Figure S1 and Figure S2 respectively show the averaged participant sensitivity and specificity for the detection of 7 urgent or malignant findings, as well as for any abnormality in exams. The averaged sensitivities and specificities are shown for the reporting round that did not use red dot® ("Without AI") and for the reporting round that used red dot® ("With AI").

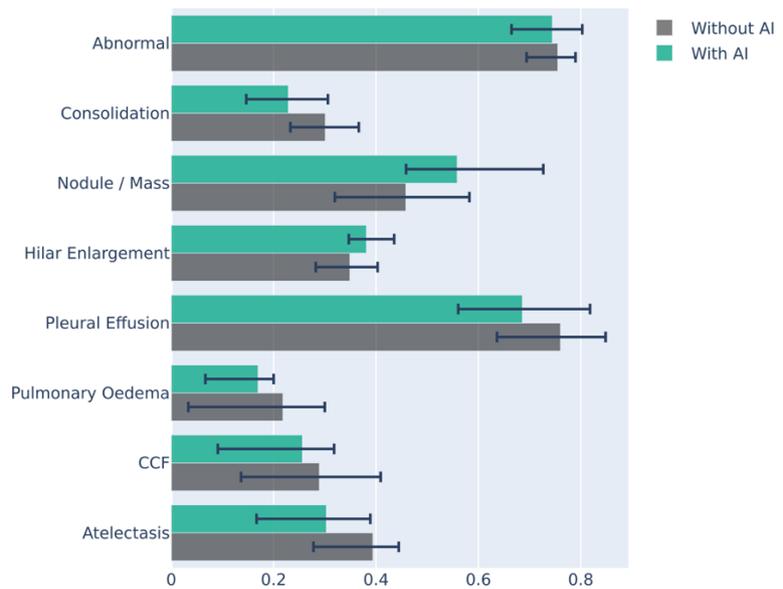

*Figure S1 - Averaged sensitivity for all 11 participants on findings. The black bars depict the 1st and the 3rd sensitivity quartiles.*

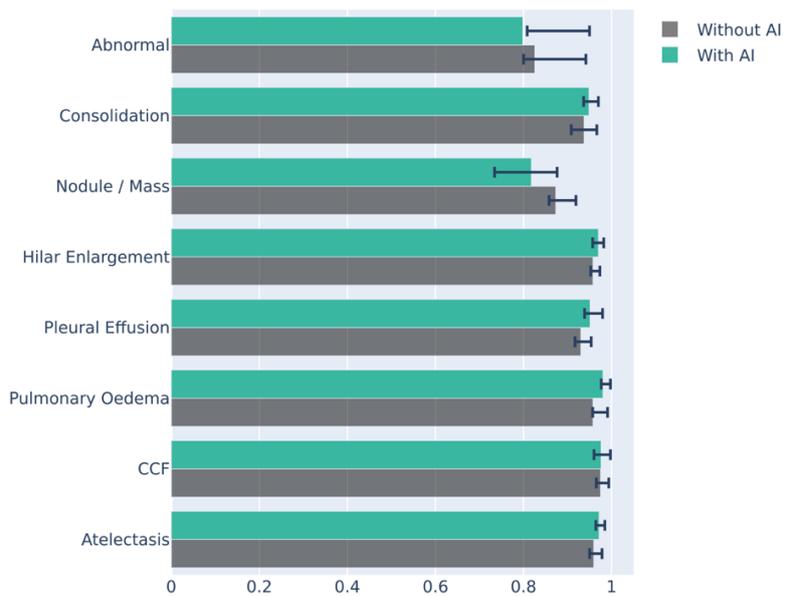

*Figure S2 - Averaged specificity for all 11 participants on findings. The black bars depict the 1st and the 3rd specificity quartiles.*

The detection rate or sensitivity is increased with AI for Nodules / Masses and Hilar Enlargements (respectively +0.10 and +0.03), which are the findings most associated with a high risk of a lung tumour diagnosis. This increase is consistent with the red dot® design, which aims at detecting with a high sensitivity Nodules, Masses and Hilar Enlargements. The sensitivity decreases with the use of AI for all other findings presented in Figure S1. Notably, the detection rates of those findings present with a high intra and inter-variability, as shown by the black ticks that represent the 1st and the 3rd quartiles for sensitivity amongst the 11 participants.

Figure S2 shows that the average specificity only decreases for the "Abnormal" and "Nodule / Mass" findings (respectively -0.03 and -0.05). This decrease is higher for exams that are identified as SLC by the AI algorithm (respectively -0.30 and -0.21) than for exams that are not identified as SLC by the AI (respectively -0.01 and +0.01). This outlines a well-known trade-off: participants tend to both generate more false-positive findings and more true-positive findings on exams that have been deemed suspicious for lung cancer by the algorithm.

However, sensitivity and specificity changes for all abnormal findings were found to be non-significant. This suggests that the AI output did not alter the clinicians' performance in detecting abnormalities on CXRs.

## II. Confidence Scores

| PARTICIPANT | CONFIDENCE WITHOUT AI | CONFIDENCE WITH AI | CHANGE IN CONFIDENCE | P-VALUE | SIGNFICANT CHANGE? | PARTICIPANT EXPERTISE |
|---|---|---|---|---|---|---|
| **ALL** | 3.54 | 3.54 | 0.00 | n/a | n/a | n/a |
| **1** | 3.11 | 3.18 | 0.07 | 1.00E+00 | No | trainee |
| **2** | 3.37 | 3.58 | 0.21 | 9.99E-05 | **Yes** | trainee |
| **3** | 3.69 | 3.05 | -0.64 | 8.33E-12 | **Yes** | consultant |
| **4** | 3.00 | 3.14 | 0.14 | 8.11E-01 | No | radiologist |
| **5** | 3.43 | 3.65 | 0.22 | 9.01E-04 | **Yes** | radiographer |
| **6** | 4.41 | 4.01 | -0.40 | 5.64E-06 | **Yes** | consultant |
| **7** | 3.27 | 2.96 | -0.31 | 4.77E-14 | **Yes** | radiographer |
| **8** | 2.85 | 3.21 | 0.36 | 1.44E-05 | **Yes** | radiologist |
| **9** | 3.56 | 3.81 | 0.25 | 9.94E-02 | No | radiographer |
| **10** | 4.25 | 4.28 | 0.03 | 1.00E+00 | No | consultant |
| **11** | 3.99 | 4.08 | 0.09 | 5.13E-01 | No | radiographer |

*Table S1 - Confidence scores for participants to the study. A p-value lower than 0.05 indicates that the change in confidence is significant. Participants that are radiologist but not consultant have a DNB/MD.*

Table S1 shows the confidence score of the 11 participants, for the reports that were produced without and with the AI. The confidence scores given by 6 of the 11 participants change significantly with the use of AI. On average, with the use of AI, 3 of those participants have had an increased confidence and 3 have had a decreased confidence in their reports. Therefore, it shows that reporters' confidence levels can vary in different ways with the use of AI, despite not being biased towards a unique direction.